\UseRawInputEncoding
\documentclass[12pt,preprint]{aastex}
\usepackage{amsmath}

\shorttitle{Probing MW by RRLs}
\shortauthors{Ablimit at el.}

\begin{document}


\title{The Milky Way Revealed by Variable Stars I: Sample Selection of RR Lyrae stars, and Evidence for the Merger History}
\author{Iminhaji Ablimit\altaffilmark{1,2}, Gang Zhao\altaffilmark{1}, Uy. Teklimakan\altaffilmark{3}, Jian-Rong Shi\altaffilmark{1},
and Kunduz Abdusalam\altaffilmark{3}}
\altaffiltext{}{For the data or other comments, it should be addressed to G. Zhao and I. Ablimit. Emails: gzhao@nao.cas.cn; iminhaji@nao.cas.cn}
\altaffiltext{1}{CAS Key Laboratory for Optical Astronomy, National Astronomical Observatories, Chinese Academy of Sciences, Beijing 100012,
China. }
\altaffiltext{2}{Department of Astronomy, Kyoto University, Kitashirakawa-Oiwake-cho, Sakyo-ku, Kyoto 606-8502, Japan.}

\altaffiltext{3}{Department of Physics, Xinjiang University, Urumqi 830046, China}


\begin{abstract}

In order to study the Milky Way, RR Lyrae (RRL) variable stars identified by Gaia, ASAS-SN and
ZTF sky survey projects have been analyzed as tracers in this work. Photometric and spectroscopic information of 3417
RRLs including proper motions, radial velocity and metallcity are obtained from observational
data of Gaia, LAMOST, GALAH, APOGEE and RAVE. Precise distances of RRLs with typical
uncertainties less than 3\% are derived by using a recent comprehensive period-luminosity-metallicity relation.
Our results from kinematical and chemical analysis provide important clues for the assembly history of the Milky Way, especially for
the Gaia-Sausage ancient merger. The kinematical and chemical trends found in this work are consistent with that of recent simulations
which indicated that the Gaia-Sausage merger is the dual origin of the Galactic thick disc and halo.
As recent similar works have found, the halo RRLs sample in this work
contains a subset
of radially biased orbits besides a more isotropic component.
This higher orbital anisotropy component amounts to $\beta\simeq 0.8$, and
it contributes between 42\% and 83\% of the halo RRLs at $4 < R(\rm kpc)<20$.

\end{abstract}


\section{Introduction}

In the past decades, tremendous efforts have been made to map and understand
the Milky Way including its assembly history and structures, and studies to date
(e.g., Searle \& Zinn 1978; Flynn et al. 1996;
Helmi et al. 1999;  Flynn et al. 2006; Nusser 2009; Bovy et al. 2012; Ness
et al. 2013; Wegg \& Gerhard 2013; Bensby et al. 2013; Rix \& Bovy 2013; Reid et al. 2014; Bird \& Flynn 2015; Gaia Collaboration et al. 2016;
Bonaca et al. 2017; Wang et al. 2018;  Gallart et al.
2019; Koppelman et al. 2019; Mackereth \& Bovy 2020; Belokurov et al. 2020;
Ablimit et al. 2020; Oman et al. 2020; Fragkoudi et al. 2020; Bobylev et al. 2020;  Ferraro et al. 2020;
Bajkova \& Bobylev 2020; Queiroz et al. 2020; Rizzuti et al. 2021; Widmark et al. 2021; de Salas \& Widmark 2021)
show that revealing the facts of the Milky Way still needs further investigations.
Stellar tracers are crucial tools to achieve the goal of studying the
Milky Way, such as RR Lyrae variable stars (RRLs) (e.g., Preston
1959; Kinman et al. 1966; Oort \& Plaut 1975; Saha 1985; Suntzeff et al. 1991;
Watkins et al. 2009; Ablimit \& Zhao 2017; Navarro et al. 2021; Iorio \& Belokurov 2021).

Metallicity distributions, kinematics and dynamics of bulge
stars can provide important constraints for the formation history of
the bulge and Milky Way (e.g., McWilliam \& Rich 1994; Zoccali et al. 2008;
Ness et al. 2013; Freeman et al. 2013; Queiroz et al. 2020).
Peculiar features of the structural components of the Galactic
bulge/bar are found by spatial distributions of bulge stars,
such as an asymmetric box-shaped bulge (Maihara et al. 1978;
Weiland et al. 1994; Dwek et al. 1995), also called as X-shaped
structure (Saito et al. 2011; Ness et al. 2012; Nataf et al. 2015).
By using RRLs, Alcock et al. (1998) showed a barred distribution, while Wesselink (1987) and Prudil et al. (2019) did not find
such shape in their data.
Moreover, results of Kunder et al. (2016) with RRLs imply a
spheroidal and pressure-supported component in the bulge.
A number of studies have described the spatial density of bulge's objects by using a simple power-law
(e.g., Harris 1976;
Pietrukowicz et al. 2012; Nataf et al. 2013; Pietrukowicz et al. 2015; Barbuy et al. 2018).
The orbits of the vast majority of RRLs in the inner Milky
Way seem to be confined to the bulge (Prudil et al. 2019;
Kunder et al. 2020), where they have witnessed the earliest
formation history of our Galaxy. There are no RRLs categorized as bulge stars in the sample used in this study,
thus the topic of the bulge is out of the scope of this work.


Most stars reside in the disc which is a highly non-axisymmetric structure still not completely
characterized, its formation and evolution being a non-solved subject.
It is widely accepted that the disc has two components according to their stellar population: the thick and thin discs (e.g., Gilmore \& Reid 1983).
Referred to the position of the Sun, the disc can be divided into the inner and outer disc: the inner disc (between the Sun
and the Galactic Centre) is composed of the more metal-rich disc stars, while the outer disc
consists of relatively metal-poor ones (e.g. Bensby et al. 2013; Anders et al. 2014; Haywood
et al. 2015). The thin disc is also categorized as inner and outer thin discs due to different
metallicities (e.g. Haywood et al. 2013; Hayden et al. 2015; Bland-Hawthorn et al. 2019).
Distributions of chemical abundance and velocity with radius have indicated that the thick
disc has a lower metallicity (e.g. Nidever et al. 2014; Hayden et al. 2015; Queiroz et al. 2020;
Ciuca et al. 2020) and slower rotation (e.g. Soubiran et al. 2003; Kordopatis et al. 2013;
Robin et al. 2017). The structure of the Milky Way thick disc have been characterized by RRLs as
tracers (e.g., Layden 1995; Amrose \& Mckay 2001). Zinn et al. (2020) and Prudil
et al. (2020) combined RRL with available spectroscopy and
Gaia DR2 astrometry to confirm the existence of metal-rich RRL
stars with the orbital properties typical of the Galactic thin disc.

The accretion and merging history of the Milky Way formation
has been studied with stellar members of the Milky Way halo
(Searle \& Zinn 1978; White \& Rees 1978;
Blumenthal et al. 1984;  Yanny et al. 2000; Newberg et al. 2002; Bullock \& Johnston 2005; Springel
et al. 2006; Bonaca et al. 2012). The halo contains about 1\% of the Galaxy's total
stellar mass, it has several components such as inner halo, outer halo, streams and numerous overdensities
(e.g., Carollo
et al. 2007; Grillmair \& Dionatos 2006; Belokurov et al. 2006; Tissera et al. 2014; Bernard et al. 2016).
Mapping the kinematics, metallicity and spatial distribution of the halo provides valuable
information that reveals the nature of the Galactic halo (e.g., Carollo
et al. 2007; Sesar et al. 2013; Helmi et al. 2017; Koppelman et al. 2018; Iorio \&
Belokurov 2019; Simion et al. 2019). More precise and new features
of the Milky Way halo have been discovered by recent survey data, and RRLs are an important population to probe the Galactic halo
(e.g. Vivas et al. 2001;  Morrison et al. 2009; Watkins et al. 2009; Mateu et al. 2018; Ablimit \& Zhao 2018; Hernitschek et al.
2018; Utkin et al. 2018; Iorio et al. 2018; Wegg et al. 2019; Iorio \& Belokurov 2019; Koposov et al. 2019; Torrealba et al.
2019). Recently, an ancient merger has been constrained with Gaia DR2,
called the Gaia-Sausage (GS; Belokurov et al. 2018) or Gaia-Enceladus
(Helmi et al. 2018),
whose debris resides in the local stellar halo. The metal-poor halo population is dominated by the GS debris,
and these stars have a triaxial 3D shape (see Iorio \& Belokurov 2019) and a
characteristic density break around 20-30 kpc from the centre of
the Galaxy (see Deason et al. 2018; Ablimit \& Zhao 2018). Iorio \& Belokurov (2021) studied the chemo-kinematics of the Gaia RR Lyrae stars
with photometrically determined metallicity, positions and proper motions (5-dimensions),
and they found that the Gaia-Sausage could contribute
50\% - 80\% of the halo RR Lyrae between galactocentric distances of 5 and 25 kpc.

As introduced above, RRLs have served as true tracers for uncovering all components,
and their estimated distances with $<$ 5\% errors derived from well-established
period-luminosity or period-luminosity-metallicity relation makes them even more reliable in
that sense (e.g., Madore \& Freedman 2012; Catelan \& Smith 2015; Neeley et al.
2017; Sesar et al. 2017).
RR Lyraes are pulsating, low-metallicity, core helium-burning
horizontal branch giants (A2 - F6
stars) with periods between 0.2 - 1.1 days and age $>$10 Gyr (Walker
1989; Smith 1995).
Recently, the number of observed RRL stars in the Galactic bulge,
disc, and halo are significantly increasing with large photometric surveys such
as the Catalina Sky Survey (Drake et al. 2014), the All-Sky Automated Survey for Supernovae (Jayasinghe et al.
2018), the Panoramic Survey Telescope and Rapid Response System (Sesar et al. 2017), the European Space Agency mission Gaia (Clementini et al. 2019), the
Optical Gravitational Lensing Experiment (Soszy$\acute{\rm n}$ski
et al. 2019), and the VISTA Variables in the V$\acute{\rm i}$a L$\acute{\rm a}$ctea survey
(D$\acute{\rm e}$k$\acute{\rm a}$ny et al. 2018).
In this work, we have analyzed RRLs detected by recent survey projects,
and the data set of the RRL sample with the spectroscopic information including metallicity and radial velocity in this work are introduced in \S 2.
The details of the distance estimation based on the new period-luminosity-metallicity
relation at mid-infrared wavelengths are also shown in \S 2.
The characterizations of the disc and halo from 7D information of RRLs along with the evidence for the merger are presented and discussed
in \S 3, \S 4 and \S 5, respectively. The paper is closed with the conclusion in \S 6.


\section{Data}
\label{sec:model}

The majority of known RRLs are of the ab-type (RRab) which pulsate in the
fundamental mode and have high amplitudes and a characteristic tooth-shaped light curve, and
of the c-type which pulsate in the radial first overtone and have more sinusoidal light
curves with lower amplitudes. From the cross-matching spectroscopic survey data, we find that the fraction of d-type RRL stars which
pulsate in both the fundamental mode and the first overtone is less than 1\% in this work. Thus, we focus on RRab and RRc stars identified from
the All-Sky Automated Survey for Supernovae (ASAS-SN) Variable stars catalog
(Jayasinghe et al. 2018), the European Space Agency mission Gaia (Gaia Collaboration 2016,
2018; Holl et al. 2018) including the specific object studies of variable stars with Gaia data release (DR) 2 (see Clementini et al. {2019} for the catalog of RRLs with Gaia DR2),
and the Zwicky Transient Facility (ZTF) catalog (Chen et al. 2020).
We made a cross-match (within an angular distance of 1 arcsec) of all the RRLs from different catalogs in order to remove multiple entries,
and we selected RRLs which have mid-infrared ($W1,W2,W3$ and $W4$ bands) magnitudes from the All WISE catalogue (e.g., Wright et al. 2010).
In order to derive spectroscopically determined metallicity and radial velocity,
we have cross-matched the RRL sample with the Large sky Area Multi-Object fiber Spectroscopic
Telescope (LAMOST) DR7 (e.g., Zhao et al. 2006, 2012;  Wang et al. 2020),
the Apache Point Observatory Galactic Evolution
Experiment (APOGEE) DR16 (e.g., Majewski
et al. 2017), the RAdial Velocity Experiment (RAVE) DR5 (e.g., Steinmetz et al.
2018) and the GALAH survey DR2 (e.g., De Silva
et al. 2015).

In total, we have 3417 RRLs with mid-infrared magnitudes, metallicity, radial velocity and proper motions.
85\% of them are RRab stars, and 15\% are RRc stars.
In our sample, the fraction of RRLs identified from Gaia, ASAS-SN and ZTF (see Figure 1) are $\sim$53\%, $\sim$33\% and $\sim$14\%, respectively.
Contributions from different surveys to spectroscopically determined metallicity and radial velocity in our sample are:
$\sim$79\% from low-resolution spectra of LAMOST DR7 (here we selected the data with signal-to-noise ratio $>$ 10), $\sim$8\% from medium-resolution spectra of LAMOST DR7, $\sim$1\% from medium-resolution spectra of RAVE DR5,
$\sim$8\% from high-resolution spectra of GALAH DR2, and $\sim$4\% from high-resolution spectra of APOGEE DR16. We also cross-matched with SDSS data (Aguado et al. 2019) but the
information was very scarce for this work. The systematic uncertainties of these spectroscopic survey data are also taken into consideration, for example the radial velocity derived by the low-resolution spectra of LAMOST DR7 has a systematic uncertainty of few $\rm km\,s^{-1}$, while the medium and high-resolution spectra of other telescopes give less systematic uncertainties. The photometric proper motions for RRLs in the work are all obtained from Gaia DR2.
Figure 2 shows distributions of observed parameters, their uncertainties,
and typical uncertainties with dashed lines.
About 35\% RRLs have both the photometrically derived metallicities (Clementini et al. 2019) and spectroscopically determined metallicities in our sample,
Figure 3 compares both metallicities for this subset of stars, for which
clear differences are visible. Considering uncertainties (which may have influence on the distance calculation) including systematic difference in spectroscopic- VS.
photometric-metallicity,
the spectroscopic metallicities are more precise (see Figure 3), thus we take spectroscopic ones instead
of the photometric ones.

After assembling the basic data set, we calculate the absolute magnitudes
by using the new theoretical period-luminosity-metallicity relations with WISE bands magnitudes. (Neeley et al. 2017).
The relations for RRab stars are (see Table 3 of Neeley et al. 2017),

\begin{subequations}\label{subequation}
\begin{align}
M_{W1}=-0.784-2.274\times{{\rm log}P} + 0.183\times{[\rm Fe/H]}\label{sub1}\\
M_{W2}=-0.774-2.261\times{{\rm log}P} + 0.190\times{[\rm Fe/H]}\\
M_{W3}=-0.800-2.292\times{{\rm log}P} + 0.188\times{[\rm Fe/H]}\\
M_{W4}=-0.799-2.298\times{{\rm log}P} + 0.196\times{[\rm Fe/H]}
\end{align}
\end{subequations}

and for RRc stars,

\begin{subequations}\label{subequation}
\begin{align}
M_{W1}=-1.341-2.716\times{{\rm log}P} + 0.152\times{[\rm Fe/H]} \\
M_{W2}=-1.348-2.720\times{{\rm log}P} + 0.153\times{[\rm Fe/H]}\\
M_{W3}=-1.357-2.731\times{{\rm log}P} + 0.157\times{[\rm Fe/H]}\\
M_{W4}=-1.355-2.735\times{{\rm log}P} + 0.166\times{[\rm Fe/H]},
\end{align}
\end{subequations}
where $P$ and $[\rm Fe/H]$ are the pulsation periods and metallicities of RRLs.
Results of Neeley et al. (2017) have showed that the typical dispersion of
the absolute magnitude is $\sim$0.02 mag from this new relations at mid-infrared wavelengths.
They also demonstrated the potential of
RRLs to be high-precision distance indicators based on these relations at
MIR wavelengths where the effects of extinction and intrinsic
dispersion are smaller.


Based on the magnitudes ($V_{\rm W}$) and absolute magnitudes ($M_{\rm W}$), we calculated
heliocentric distances ($D_{\rm h}$) by using,

\begin{equation}
D_{\rm h} = 10^{(V_{\rm W}-M_{\rm W}-10)/5} {\rm kpc}.
\end{equation}

For the distance of each star, we take the averaged value from four distances
derived by using four magnitudes and absolute magnitudes. There is a few percent typical distance dispersion in the
average distance computed. The cumulative distribution function for $D_{\rm h}$ is given in Figure 4,
most of the stars in our sample have heliocentric distance $<15$ kpc.
Recently, it has been discussed that
distances derived from mid-infrared period-luminosity relations are more accurate than distances obtained
from parallaxes (e.g., Sesar et al. 2017).
Considering all possible uncertainties in the distance calculation with the new theoretical period-luminosity-metallicity relations, the typical uncertainty of the estimated
 distance can be lower than 3\% in this work.
The 3D positions of RRLs, the projection of galactocentric
distance on the Galactic plane ($R$), and
galactocentric distance ($r$) in the Cartesian coordinate system are calculated as

\begin{subequations}\label{subequation}
\begin{align}
x = {\rm R}_0 - D_{\rm h}{\rm cos}\,b\, {\rm cos}\,l \\
y = D_{\rm h}{\rm cos}\,b\, {\rm sin}\,l\\
z = D_{\rm h}{\rm sin}\,b\\
R = \sqrt{x^2 + y^2}\\
r = \sqrt{x^2 + y^2 + z^2},
\end{align}
\end{subequations}
where ${\rm R}_0$ is the distance from the Sun to the Galactic center; the recent
most accurate value, $8.122\pm0.031$ kpc (GRAVITY collaboration et al. 2018), is adopted in this work. $l$ and $b$ are
Galactic longitude and latitude (see Figure 1 for the distributions of $l$ and $b$ of RRLs), respectively.
In the following sections we will characterize the Galactic disc and halo
by analyzing 7D information, i.e. metallicities (see Figure 4 for the metallicity distribution in the r and z plane),
3D positions and 3D velocities, of RRLs.  Note that the heliocentric distance, the observed Galactic coordinates and velocities
are used to obtain the Galactocentric Cartesian, cylindrical and spherical
coordinates taking into account the errors on the Galactic parameters (i.e. origin of galactic coordinates,
solar radius, and solar azimuthal velocity with respect to the galactic
center) by using the conventions in astropy (Astropy Collaboration et al. 2018).


\section{The Chemo-Kinematic Clues for the Assembly History of the Galaxy}

Mergers of satellite galaxies have been suggested to have the crucial role to form the galaxy
in the $\Lambda$CDM cosmology (e.g., White \& Rees 1978).
Mergers can significantly influence the stellar population in a galaxy,
therefore motions (i.e. rotational velocity) and chemical (i.e. metallicity)
properties of stars can provide
important clues to understand the formation history of the galaxy.
Figure 5 shows the relation of the azimuthal (in the direction of the Galactic rotation) velocity ($V_\phi$)
and metallicity (Fe/H).
The area bounded in magenta in Figure 5 is populated by stars with $[\rm Fe/H] > -0.8$ and higher azimuthal
velocity ($V_\phi > 200 \,\rm km\,s^{-1}$), which belongs to the thin disc. The
area bounded in green
corresponds to the thick disc population with intermediate metallicity of
$[\rm Fe/H] \sim -1.2$ and
azimuthal velocity of $80 <V_\phi < 200 \,\rm km\,s^{-1}$. There is also a
stellar population corresponding to the
stellar halo which includes the GS structure (see Belokurov et al. 2018) at
lower azimuthal velocity ($V_\phi \sim 0\, \rm km\,s^{-1}$) and lower
metallicity ($-2.3 < [\rm Fe/H] < -1.1$)(orange bounded area of the figure).
The negative and inverted $V_\phi$-metallicity gradients (magenta and green dashed lines in Figure 5) are also shown up in the thin and
thick disc populations (Figure 5), and they may be caused by the asymmetric drift and inside-out disc formation (e.g., Sch$\ddot{\rm o}$nrich \&
Mcmillan 2017; Kawata et al. 2018; Minchev et al. 2019; Belokurov et al. 2020). A negative radial metallicity gradient and
epicycle motions of stars (e.g. Allende et al. 2016) can explain the reason why the young thin disc occupies the high-rotation velocity and
metal-rich part of the space, and its metallicity-rotation velocity relation.
The population with $V_\phi < 0 \,\rm km\,s^{-1}$ is consistent with the Splash
structure identified in the halo
by previous studies (see Belokurov et al. 2020). In Figure 6, we plot
$V_\phi$ vs $|z|$ (top panels) and $V_\phi$ vs R (bottom panels) for the
metal-rich $[\rm Fe_H]>-0.7 $(left panels) and metal-poor $[\rm Fe/H]<-0.7$ (right
panels) RRL stars.
In the metal-poor population, $V_\phi <0$ stars are within the continuous
extension of the whole dataset
but in the metal-rich sample the $V_\phi <0$ stars look a bit detached from
the general overall distribution
of the data points, precisely at the location of the Solar neighborhood,
where the Splash is located. It is worth noting that there are strong selection effects which are yielded by each
survey's footprint and limited data etc., and selection bias may influence our results.
If selection effects could be neglected, it can be concluded that the
Splash population is considerably
smaller than that of the halo.

Our results are consistent with the trend and schematic picture between azimuthal (rotation) velocity and metallicity
presented by Belokurov et al. (2020), and provides comprehensive clues for the formation history of
the Galaxy inferred from the simulation results of Grand et al. (2020).
The relation of azimuthal (Galactic rotation) velocity and metallicity in Figure 5 can be explained by the
cosmological magnetohydrodynamic simulation results of Grand et al. (2020).
Their simulation results indicated that the proto-disc evolves and self-enriches,
while the proto-halo is formed before the GS merger. The starburst-disc forms quickly in a more centrally bound region (thin
disc) which carries more rotation and chemically enriched stars.
Then, multiple smaller mergers dynamically heat the older and metal-poor
proto-disc stars more than the younger and metal-rich ones, giving rise to the relation
between metallicity and azimuthal velocity. The inner region of the halo contains
rotationally-supported and relatively metal-rich stars which were dynamically pushed out from
the proto-disc by the GS merger into the (Splash and) halo, while the outer
halo region is dominated by the metal-poor proto-halo (Grand et al. 2020). They also infer that the
stars formed in the bridge of cold gas immediately preceding the merger may contribute
to this in-situ, metal-rich halo, highlighting the complexity involved in the genesis of
these components.

\section{Properties of Thin Disc, Thick Disc and Halo}

Figure 7 shows the spatial distributions of RRLs in our sample of 3417 RRL stars. We separate the non-rotating halo from the disc component with a
high azimuthal velocity, and our average values of azimuthal velocity or metallicity which are set up for these two components are similar with the values given in Table 1. The halo stars ($\sim$12.9\% of our sample) are distributed
beyond $|\rm z| = 5$ kpc, while thick disc ($\sim$64\% of the sample) and thin disc ($\sim$16.4\% of the sample) RRLs are located at $0.5\leq|\rm z| < 5$ kpc
and $|\rm z| < 0.5$ kpc, respectively. Disc and halo RRLs are treated in the cylindrical ($r,\,\phi,\,z$) and spherical ($r,\,\phi,\,\theta$) coordinates , respectively.  6.7\% of the RRLs are unclassified, their vertical positions and azimuthal velocities differ significantly
from the typical
properties of disc and halo, thus, they were excluded from the following
analysis.
We do not study the bulge in this work, because almost all of the sample
has R $>$ 3 kpc.
Based on the following the total
likelihood function,  we can apply the Gaussian Model in order to derive the best values of the
rotational and chemical characteristics of different components (Figure 8).
By fitting the velocities ($\textbf{V}$) and metallicity ($[\rm Fe/H]_{\rm i}$) distributions with errors,
the likelihood of observed a star (i) can be simply written as
$\mathcal{L} = \mathcal{N}(\textbf{V}, \Sigma_{\rm i}) \mathcal{N}([\rm Fe/H])$, where $\Sigma_{\rm i}$ is the covariance matrix or velocity
dispersion tensor. With this method which is the product of the likelihoods of all stars
in a given Galactic volume bin (the volume being sampled by the
data), we fit the different components to derive the best values of the azimuthal velocity and metallicity by adopting a Markov Chain Monte
Carlo (MCMC) Ensemble sampler in the fitting process (Foreman-Mackey et al. 2013). In the MCMC sample process (Emcee), 200
walkers and 800 steps are used as burn-in, then followed by 2000
steps to get the posterior distributions for each bin.


Figure 9 shows our fitting results of the azimuthal velocity and metallicity for different components,
and Table 1 summarizes the best estimated parameters. The mean azimuthal velocities and mean metallicities in Table 1
are the typical values of the galactic rotation and metallicity of the thin disc, thick disc and halo components.
The mean azimuthal velocity of the thin disc indicates that the motion of the Sun is around $232.5\,\rm km\,s^{-1}$, which
is consistent with the recent value derived from classical cepheids (Ablimit et al. 2020). The faster azimuthal velocity and more metal-rich
features of the thin disc compared to other components are consistent with previous studies (e.g., Preston 1959; Taam et al. 1976; Layden 1995; Mateu \& Vivas 2018; Marsakov et al. 2019; Prudil et al. 2020; Zinn et al. 2020; Iorio \& Belokurov 2021). There are relatively more thick disc stars in this work, and they do not rotate as fast as the thin disc stars.
Thick disc has an azimuthal mean velocity
in between the thin disc and the halo, and it is only slightly more metal-rich
than the halo (see Table 1). Majority of them are distributed within $r=14$ kpc, which is consistent with the
theories of the thick disc formation and recent finding (e.g., Iorio \& Belokurov 2021). It has been widely accepted that the kinematics of
disc stars (compared to the halo) is easily affected by non-axisymmetric structures such as the spiral arms, the bar, the giant molecular clouds,
and in-falling dark matter sub-structure etc., thus it makes their motions unstable with time.

As shown in Table 1, the mean azimuthal velocity of the halo stars is up to 57.8 $\rm km\,s^{-1}$ (non-zero),
in agreement with recent studies (e.g., Deason et al. 2017; Belokurov et al. 2018b;  Wegg
et al. 2019; Iorio \& Belokurov 2021),
thus indicating the contribution of the GS substructure to the halo. Indeed, it has been pointed out that the inner halo is clearly influenced by the
the Gaia Sausage event (see
e.g. Belokurov et al. 2018b; Myeong et al. 2018). The GS merger has contributed to the rotation (azimuthal velocity)
of the halo by pushing the more rotationally supported proto-disc
stars out into the proto-halo (see above section for more discussion),  and a bimodal
signature in the radial velocity space has been caused by the collision between the progenitor of the GS
and the proto-Galaxy. A model has been used to describe the mixture of two components in the
halo (see Lancaster et al. (2019) and Necib et al. (2019)),
\begin{equation}
\mathcal{L}_{\rm M} = \sum_{\rm j} f_{\rm j}\mathcal{L}_{\rm j},
\end{equation}
where the component weights $f_{\rm j}$ sum up to 1, and $\mathcal{L}_{\rm j}$ is each component and discussed as (see $\mathcal{L}$) above.
The prior distributions of the anisotropic component
reflect our knowledge of the radially anisotropic nature of the halo. Indeed, the radial velocities of halo RRLs in this work (Figure 10) are
also described better by the the Gaussian
Mixture Model. In the next section, we present the anisotropy parameter study in order to see the contribution of the radially biased (anisotropic) component.

\section{Contribution of GS Component to the Halo}

After using the double-component fit (see section 4) to distinguish
two components of the halo, from the data we calculate the kinematic
properties to examine contribution of the GS component with its higher azimuthal velocity with
respect to the non-rotating halo.
The kinematic properties of the stellar halo can be
summarized well by the anisotropy parameter $\beta$,

\begin{equation}
\beta = 1 - \frac{\sigma^2_\theta + \sigma^2_\phi}{2\sigma^2_{\rm r}}.
\end{equation}
where,
\begin{subequations}\label{subequation}
\begin{align}
\sigma^2_{\rm r} = \overline{V^2_{{\rm r}}} - {\overline{V_{{\rm r}}}}^2\\
\sigma^2_\theta = \overline{V^2_{\theta}} - {\overline{V_{\theta}}}^2\\
\sigma^2_\phi = \overline{V^2_{\phi}} - {\overline{V_{\phi}}}^2
\end{align}
\end{subequations}
$\beta$ quantifies the degree of velocity anisotropy of the
stellar orbits system: radially biased ($\beta > 0$, perfectly
radial $\beta = 1$), perfectly isotropic ($\beta = 0$), tangentially
biased ($\beta < 0$, perfectly circular $\beta = -\infty$).

Upper panel of Figure 11 shows the resulting anisotropy parameters of the two stellar halo components as a function of galactocentric distance $r$.
The radially biased GS component is more anisotropic ($\beta\sim0.8$),
and it has relatively metal rich ($[\rm Fe/H]>-1.6$ dex) stars. The fraction of the GS stars goes from 42\% to 83\% (here only considered mean values)
of the halo RR Lyrae at $4 < R(\rm kpc)<20$
(see lower panel of Figure 11). Thus,  the anisotropic component has dominant
contributions in this particular data set, and this is in good agreement with
recent studies (e.g., Iorio \& Belokurov 2021; Naidu et al. 2021). The another
halo component is more isotropic ($\beta\sim0.42$) with relatively metal poor
stars ($[\rm Fe/H]<-1.6$ dex). We confirm that the halo can be described with
two populations with different properties (see e.g. Chiba \& Beers 2000;
Carollo et al. 2010;  Belokurov et al. 2018b; Iorio \& Belokurov 2021).

\section{Conclusion}

We have analyzed several hundreds of thousands of RRLs detected by Gaia, ASAS and ZTF surveys, and
gathered a sample of 3417 RRLs which have proper motions, radial velocity and metallicity
from Gaia, LAMOST, GALAH, APOGEE and RAVE data. A recent comprehensive period-luminosity-metallicity
relation is used to estimate distances, with very small uncertainties
(typically $< 3\%$).
The reliable distances enable us to map the chemo-kinematics of the RRLs to
identify the RRL populations with distinct properties.

By using precise 7-dimensional data (3D positions, 3D velocities and metallicity) of RRLs for the first time,
we provide kinematical and chemical proofs for the assembly history of the Milky Way.
Comparing with cosmological magnetohydrodynamic simulations of the formation of Milky Way-mass galaxies,
the metallicity and azimuthal velocity trends in this work are consistent with those in the simulated
results from the formation scenario including the proto-/starburst-disc, proto-halo and mergers (Grand et al. 2020),
and our results support the indications of Grand et al. (2020) about the dual origin of the Galactic thick disc and halo from GS merger.
Observational properties of the Galaxy main components obtained
in this work demonstrate that the gas-rich proto-disc evolves and becomes
heavier and more
enriched as time proceeds, metal-rich and fast-rotating stars can be born
in the thin disc. The GS merger and starburst materials generates relatively metal-poor and medium
rotating stars of thick disc, the relatively large-dynamical impact of the GS merger scatters more
metal-rich stars into the halo, and repeated mergers may move the older/metal-poorer populations to the halo-like orbits and create the Splash (see also Grand et al. 2020).

The halo RRLs in the sample can be described by a radially supported component
plus another one more isotropic component.
We also find that the radially biased component is a part with the high anisotropy, and it contributes to about 42\% - 83\%
of the halo stars at $4 < R(\rm kpc)<20$. Our results further support that the inner Galactic stellar halo
is dominated by the major GS merger event.
In the future work, this RRL
sample can be used to explore the new substructures of the Milky Way
and Galactic potential.
We also expect that future data releases from large sky survey projects may make it
possible to collect more RRLs with the precise distance estimation and other needed
observational information for deriving new constraints on the nature of the Milky
Way including the bulge and outer halo.

\begin{acknowledgements}

We thank the referee for their useful comments that improve the paper.
This study is supported by the National Natural Science
Foundation of China under grant Nos. 11988101, 11890694,
and National Key R\&D Program of China No. 2019YFA0405502.

This work use the data of ASAS-SN which is supported by the Gordon and Betty Moore
Foundation through grant GBMF5490 to the Ohio State University,
and NSF grants AST-1515927 and AST-1908570. Development
of ASAS-SN has been supported by NSF grant AST-0908816, the
Mt. Cuba Astronomical Foundation, the Center for Cosmology and
AstroParticle Physics at the Ohio State University, the Chinese
Academy of Sciences South America Center for Astronomy (CASSACA), the Villum Foundation, and George Skestos.

This publication made use of data from the European Space Agency mission Gaia (https://www.cosmos.esa.int/ gaia),
processed by the Gaia Data Processing and Analysis Consortium (DPAC, https://www.cosmos.esa.int/web/gaia/ dpac/consortium).
Funding for the DPAC has been provided by national institutions, in particular the institutions participating in the Gaia Multilateral Agreement.

Based on observations obtained with the Samuel Oschin
Telescope 48 inch and the 60 inch Telescope at the Palomar
Observatory as part of the Zwicky Transient Facility project.
ZTF is supported by the National Science Foundation under
grant No. AST-1440341 and a collaboration including Caltech,
IPAC, the Weizmann Institute for Science, the Oskar Klein
Center at Stockholm University, the University of Maryland,
the University of Washington, Deutsches Elektronen-Synchrotron and
Humboldt University, Los Alamos National Laboratories, the TANGO Consortium of Taiwan, the University of
Wisconsin at Milwaukee, and Lawrence Berkeley National
Laboratories. Operations are conducted by COO, IPAC, and
UW. SED Machine is based upon work supported by the
National Science Foundation under grant No. 1106171.

This work has used the data products from the Wide field Infrared Survey Explorer (WISE), which is a joint project of the University of California,
Los Angeles, and the Jet Propulsion Laboratory/California Institute of Technology, funded by the National Aeronautics and Space Administration.

The work also have used the data from the Large Sky Area Multi-Object Fiber Spectroscopic Telescope (LAMOST) which is a National Major
Scientific Project built by the Chinese Academy of Sciences. Funding for the project has been provided by the National Development and
Reform Commission. LAMOST is operated and managed by the National Astronomical Observatories, Chinese Academy of Sciences.

This work used the data from the GALAH survey which is based on observations made at the Anglo Australian Telescope,
under programmes A/2013B/13, A/2014A/25,
A/2015A/19, A/2017A/18. We acknowledge the traditional owners
of the land on which the AAT stands, the Gamilaraay people, and pay
our respects to elders past, present and emerging. This paper includes
data that have been provided by AAO Data Central (datacentral.
org.au).

Funding for the Sloan Digital Sky Survey IV has been provided by the Alfred
P. Sloan Foundation, the U.S. Department of Energy Office of Science, and the
Participating Institutions. SDSS- IV acknowledges support and resources from
the Center for High-Performance Computing at the University of Utah. The
SDSS web site is www.sdss.org. SDSS-IV is managed by the Astrophysical Research
Consortium for the Participating Institutions of the SDSS Collaboration
including the Brazilian Participation Group, the Carnegie Institution for Science,
Carnegie Mellon University, the Chilean Participation Group, the French
Participation Group, Harvard-Smithsonian Center for Astrophysics, Instituto
de Astrof¨¬sica de Canarias, The Johns Hopkins University, Kavli Institute for
the Physics and Mathematics of the Universe (IPMU) / University of Tokyo,
Lawrence Berkeley National Laboratory, Leibniz Institut f¨¹r Astrophysik
Potsdam (AIP), Max-Planck-Institut f¨¹r Astronomie (MPIA Heidelberg),
Max-Planck-Institut fur Astrophysik (MPA Garching), Max-Planck-Institut
fur Extraterrestrische Physik (MPE), National Astronomical Observatory of
China, New Mexico State University, New York University, University of
Notre Dame, Observat¨®rio Nacional / MCTI, The Ohio State University,
Pennsylvania State University, Shanghai Astronomical Observatory, United
Kingdom Participation Group, Universidad Nacional Aut¨®noma de M¨¦xico,
University of Arizona, University of Colorado Boulder, University of Oxford,
University of Portsmouth, University of Utah, University of Virginia, University
of Washington, University of Wisconsin, Vanderbilt University, and Yale
University.

 Funding
for RAVE has been provided by: the Leibniz-Institut fur Astrophysik Potsdam
(AIP); the Australian Astronomical Observatory; the Australian National University; the Australian Research Council; the French National Research Agency
(Programme National Cosmology et Galaxies (PNCG) of CNRS/INSU with INP
and IN2P3, co-funded by CEA and CNES); the German Research Foundation
(SPP 1177 and SFB 881); the European Research Council (ERC-StG 240271
Galactica); the Istituto Nazionale di Astrofisica at Padova; The Johns Hopkins
University; the National Science Foundation of the USA (AST-0908326); the
W. M. Keck foundation; the Macquarie University; the Netherlands Research
School for Astronomy; the Natural Sciences and Engineering Research Council of Canada; the Slovenian Research Agency (research core funding no.
P1-0188); the Swiss National Science Foundation; the Science \& Technology
Facilities Council of the UK; Opticon; Strasbourg Observatory; and the Universities of Basel, Groningen, Heidelberg, and Sydney. TZ acknowledges financial support of the Slovenian Research Agency (research core funding No.
P1-0188) and of the ESA project PHOTO2CHEM (C4000127986). FA is grateful for funding from the European Union¡¯s Horizon 2020 research and innovation program under the Marie Skodowska-Curie grant agreement No. 800502.

This research has made use of the VizieR catalogue access tool,
CDS, Strasbourg, France. This research also made use of ASTROPY,
a community-developed core PYTHON package for astronomy
(Astropy Collaboration et al. 2018).

\end{acknowledgements}



Ablimit, I., \& Zhao, G. 2017, ApJ, 846, 10

Ablimit, I., \& Zhao, G. 2018, ApJ, 855, 126

Ablimit I., Zhao G., Flynn C., Bird S. A., 2020, ApJ, 895, L12

Aguado, D. S., Ahumada, R., Almeida, A. et al. 2019, ApJS, 240, 23

Alcock, C., Allsman, R. A., Alves, D. R. et al., 1998, ApJ, 492, 190

Allende Prieto C., Kawata D., Cropper M., 2016, A\&A, 596, A98

Amrose, S. \& Mckay, T., 2001, 560, 151

Anders F. et al., 2014, A\&A, 564, A115

Astropy Collaboration et al., 2018, AJ, 156, 123

Bajkova, A. T. \& Bobylev, V. V., 2020 eprint arXiv:2007.02350

Barbuy, B., Chiappini, C. \& Gerhard, O., 2018, ARA\&A, 56, 223

Bensby T., et al., 2013, A\&A, 549, A147

Belokurov, V., Zucker, D. B., Evans, N. W. et al., 2006, ApJ, 642, 137

Belokurov V., Deason A. J., Koposov S. E., Catelan M., Erkal D., Drake A.
J., Evans N. W., 2018a, MNRAS, 477, 1472

Belokurov V., Erkal D., Evans N. W., Koposov S. E., Deason A. J., 2018b,
MNRAS, 478, 611

Belokurov V., Deason A. J., Erkal D., Koposov S. E., Carballo-Bello J. A.,
Smith M. C., Jethwa P., Navarrete C., 2019, MNRAS, 488, L47

Belokurov V., Sanders J. L., Fattahi A., Smith M. C., Deason
A. J., Evans N. W., Grand R. J. J., 2020, MNRAS, 494, 3880

Bernard, E. J., Ferguson, A. M. N., Schlafly, E. F., et al. 2016, MNRAS,
463, 1759

Bird, S. A. \& Flynn, C., 2015, MNRAS, 452, 2675

Bland-Hawthorn J., et al., 2019, MNRAS, 486, 1164

Blumenthal, G. R., Faber, S. M., Primack, J. R. \& Rees, M. J., 1984, Nature, 311, 517

Bobylev, V. V., Krisanova, O. I. \& Bajkova, A. T., 2020, Astronomy Letters, 46, 439

Bovy J., Rix H.-W., Liu C., Hogg D. W., Beers T. C., \& Lee Y. S.
2012, ApJ, 753, 148

Bonaca, A., Juric, M., Ivezic, Z., et al. 2012, AJ, 143, 105

Bonaca A., Conroy C., Wetzel A. et al., 2017,
ApJ, 845, 101

Bullock, J. S. \& Johnston, K. V., 2005, ApJ, 635, 931

Carollo D. et al., 2010, ApJ, 712, 692

Catelan, M. \& Smith, H. A., 2015, Book: Pulsating Stars (Wiley-VCH)

Clementini, G., Ripepi, V., Molinaro, R. et al., 2019, A\&A, 622, A60

Chen, X. D., Wang, S., Deng, L. C., de Grijs, R. et al. 2020, ApJS, 249, 18

Chiba M., Beers T. C., 2000, AJ, 119, 2843

Ciuca, I., Kawata, D., Miglio, A. et al., 2020, eprint arXiv:2003.03316

De Silva, G. M., Freeman, K. C., Bland-Hawthorn, J., et al. 2015, MNRAS,
449, 2604

Deason, A. J., Belokurov, V., Koposov, S. E., Lancaster, L., 2018, ApJ, 862,
L1

D$\acute{\rm e}$k$\acute{\rm a}$ny, I.,  Hajdu, G., Grebel, E. K. et al. 2018, ApJ, 857, 54

Drake, A. J., et al., 2014, ApJS, 213, 9

de Salas, P. F. \& Widmark, A., 2021, Reports on Progress in Physics, 84, 104901

Dwek, E., Arendt, R. G., Hauser, M. G., et al. 1995, ApJ, 445,
716

Ferraro, F. R., Pallanca, C., Lanzoni, B. et al., 2020, Nature Astronomy, 244

Flynn, C., Sommer-Larsen, J. \& Christensen, P. R., 1996, MNRAS, 281, 1027

Flynn, C., Holmberg, J., Portinari, L., et al. 2006, MNRAS, 372, 1149

Foreman-Mackey, D., Hogg, D. W., Lang, D., \& Goodman, J. 2013, PASP, 125, 306,

Fragkoudi, F., Grand, R. J. J., Pakmor, R. et al., 2020, MNRAS, 494, 5936

Freeman, K., Ness, M., Wylie-de-Boer, E. et al. 2013, MNRAS, 428, 3660

Gaia Collaboration, Prusti, T., de Bruijne, J. H. J., et al. 2016, A\&A, 595, A1

Gaia Collaboration, Abuter, R., Amorim, A., et al. 2018, A\&A, 615, L15

Gallart C., Bernard E. J., Brook C. B., Ruiz-Lara T., Cassisi S.,
Hill V., Monelli M., 2019, Nature Astronomy, 3, 932

Gilmore G. \& Reid N., 1983, MNRAS, 202, 1025

Grand, R. J. J., Kawata, D., Belokurov, V. et al. 2020, MNRAS, 497, 1603

Gravity Collaboration, Abuter, R., Amorim, A., et al. 2018, A\&A, 615, L15

Grillmair, C. J., \& Dionatos, O. 2006, ApJL, 643, L17

Harris, W. E., 1976, AJ, 81, 1095

Hayden M. R., et al., 2015, ApJ, 808, 132

Haywood, M., Di Matteo, P., Lehnert, M. D. et al., 2013, A\&A, 560, 109

Haywood, M., Di Matteo, P. \& Snaith, O. et al., 2015, A\&A, 579, 5

Helmi A., White S. D. M., de Zeeuw P. T. \& Zhao H., 1999, Nature,
402, 53

Helmi, A., Veljanoski, J., Breddels, M. A. et al., 2017, A\&A, 598, 58

Helmi, A., Babusiaux, C., Koppelman, H. H., et al. 2018,
Nature, 563, 85

Hernitschek N., et al., 2018, ApJ, 859, 31

Holl B., Audard, M., Nienartowicz, K. et al., 2018, A\&A, 618, A30

Iorio G., Belokurov V., Erkal D., Koposov S. E., Nipoti C., Fraternali F., 2018, MNRAS, 474, 2142

Iorio G. \& Belokurov V., 2019, MNRAS, 482, 3868

Iorio, G. \& Belokurov, V. 2021, MNRAS, 502, 5686

Jayasinghe, T., Kochanek, C. S., Stanek, K. Z. et al. 2018, MNRAS, 477, 3145

Kawata D. et al., 2018, MNRAS, 473, 867

Kinman T. D., Wirtanen C. A., Janes K. A., 1966, ApJS, 13, 379

Koppelman H., Helmi A., Veljanoski J., 2018, ApJ, 860, L11

Koppelman, H. H., Helmi, A., Massari, D. et al., 2019, A\&A 631, L9

Koposov S. E., et al., 2019, MNRAS, 485, 4726

Kordopatis G. et al., 2013, AJ, 146, 134

Kunder, A., Rich, R. M., Koch, A., et al. 2016, ApJL, 821, L25

Lancaster L., Koposov S. E., Belokurov V., Evans N.W., Deason A. J., 2019,
MNRAS, 486, 378

Layden A. C., 1995, AJ, 110, 2288

Mackereth, J. T. \& Bovy, J., 2020, MNRAS, 492, 3631

Madore, B. F. \& Freedman, W. L., 2012, ApJ, 744, 132

Majewski, S. R., Schiavon, R. P., Frinchaboy, P. M., et al. 2017, AJ, 154, 94

Maihara, T., Oda, N., Sugiyama, T., \& Okuda, H. 1978, PASJ,
30, 1

Marsakov V. A., Gozha M. L., Koval, V. V., 2019, Astron. Rep., 63, 203

Mateu C. \& Vivas A. K., 2018, MNRAS, 479, 211

Mateu C., Read J. I., Kawata D., 2018, MNRAS, 474, 4112

McWilliam, A., \& Rich, R. M. 1994, ApJS, 91, 749

Minchev I. et al., 2019, MNRAS, 487, 3946

Morrison H. L., et al., 2009, ApJ, 694, 130

Myeong G. C., Evans N. W., Belokurov V., Sand ers J. L., Koposov S. E.,
2018, ApJ, 863, L28

Naidu R. P., Conroy C., Bonaca A., Johnson B. D., Ting Y.-S., Caldwell N.,
Zaritsky D., Cargile P. A., 2020, ApJ, 901, 48

Nataf, D. M., Gould, A., Fouque, P. et al., 2013, ApJ, 769, 88

Nataf, D. M., Udalski, A., Skowron, J., et al. 2015, MNRAS,
447, 1535

Navarro, M. G., Minniti1, D., Capuzzo-Dolcetta, R. et al., 2021, A\&A 646, A45

Necib L., Lisanti M., Belokurov V., 2019, ApJ, 874, 3

Neeley, J. R., Marengo, M., Bono, G. et al., 2017, ApJ, 841, 84

Ness, M., Freeman, K., Athanassoula, E., et al. 2012, ApJ, 756,
22

Ness M., et al., 2013, MNRAS, 430, 836

Newberg, H. J., Yanny, B., Rockosi, C., et al., 2002, ApJ, 569, 245

Nidever, D. L., Bovy, J., Bird, J. C. et al., 2014, 796, 38

Nusser, A. 2009, ApJ, 706, 113

Oman, K. A., Brouwer, M. M., Ludlow, A. D. et al., 2020, eprint arXiv:2006.06700

Oort J. H., Plaut L., 1975, A\&A, 41, 71

Petac, M., 2020 Physical Review D, 102, 123028

Pietrukowicz, P., Udalski, A., Soszynski, I. et al. 2012, ApJ, 750, 169

Pietrukowicz, P., Koz?owski, S., Skowron, J. et al. 2015, ApJ, 811,  113

Preston G. W., 1959, ApJ, 130, 507

Prudil, Z., D$\acute{\rm e}$k$\acute{\rm a}$ny, I., Catelan M. et al., 2019, MNRAS, 484, 4833

Prudil, Z., D$\acute{\rm e}$k$\acute{\rm a}$ny, I., Grebel, E. K. \& Kunder, A., 2020, MNRAS 492, 3408

Queiroz, A. B. A., Chiappini, C., Perez-Villegas, A. et al., 2020, eprint arXiv:2007.12915

Reid, M. J., Menten, K. M., Brunthaler, A., et al. 2014, ApJ, 783,
130

Rix, H. W. \& Bovy, J., 2013, A\&ARv, 21, 61

Rizzuti, F., Cescutti, G., Matteucci, F. et al., 2021, MNRAS, 502, 2495

Robin, A. C., Bienayme, O., Fernandez-Trincado, J. G. \& Reyle, C., 2017, A\&A, 605, 1

Saha A., 1985, ApJ, 289, 310

Saito, R. K., Zoccali, M., McWilliam, A., et al. 2011, AJ, 142,
76

Sch$\ddot{\rm o}$nrich, R \& Mcmillan, P. J. 2017, MNRAS, 467, 1154

Sesar, B., Grillmair, C. J., Cohen, J. G. et al., 2013, ApJ, 776, 26

Sesar, B., Fouesneau, M., Price-Whelan, A. M. et al., 2017, ApJ, 838, 107

Searle, L., \& Zinn, R., 1978, ApJ, 225, 357

Simion I. T., Belokurov V. \& Koposov S. E., 2019, MNRAS,
482, 921

Smith, R. C., 1995, Observational Astrophysics,
Book (Cambridge, UK: Cambridge University Press), pp. 467. ISBN 0521278341

Soszynski I., Udalski, A., Wrona, M. et al., 2019, Acta Astronomica, 69, 321

Soubiran, C., Bienayme, O. \& Siebert, A., 2003, A\&A, 398, 141

Steinmetz, M., Zwitter, T., Matijevic, G. et al., 2018, RNNAS, 2, 194

Suntzeff N. B., Kinman T. D., Kraft R. P., 1991, ApJ, 367, 528

Taam R. E., Kraft R. P., Suntzeff N., 1976, ApJ, 207, 201

Tissera, P. B., Beers, T. C., Carollo, D., \& Scannapieco, C. 2014, MNRAS,
439, 3128

Torrealba G., et al., 2019, MNRAS, 488, 2743

Utkin, N. D., Dambis, A. K., Rastorguev, A. S. et al., 2018, Astronomy Letter, 44, 688

Venn K. A., Irwin M., Shetrone M. D., Tout C. A., Hill V. \& Tolstoy
E., 2004, AJ, 128, 1177

Vivas A. K., et al., 2001, ApJ, 554, L33

Walker, A. R., 1989, PASP, 101, 570

Wang, R., Luo, A-L., Chen, J.-J. et al., 2020, ApJ, 891, 23

Wang, W., Han, J., Cole, S. et al., 2018, MNRAS, 476, 5669

Watkins, L. L., Evans, N. W., Belokurov, V. et al., 2009, MNRAS, 398, 1757

Wegg C. \& Gerhard O., 2013, MNRAS, 435, 1874

Wegg C., Gerhard O., Bieth M., 2019, MNRAS, 485, 3296

Weiland, J. L., Arendt, R. G., Berriman, G. B., et al. 1994, ApJL,
425, L81

Wesselink, T. J. H., 1987,  Ph.D. thesis, Katholieke Univ. Nijmegen

Widmark, A., de Salas,  P. F. \& Monari, G., 2021, A\&A 646, A67

Wright, E. L., Eisenhardt, P. R. M., Mainzer, A. K., et al. 2010, AJ, 140, 1868

Yanny, B., Newberg, H. J., Kent, S., et al., 2000, ApJ, 540, 825

Zhao, G., Chen, Y.-Q., Shi, J.-R., et al. 2006, ChJAA, 6, 265

Zhao, G., Zhao, Y. H., Chu, Y. Q., et al. 2012, RAA, 12, 723

Zinn R., Chen X., Layden A. C., Casetti-Dinescu D. I., 2020, MNRAS, 492,
2161

Zoccali, M., Hill, V., Lecureur, A., et al. 2008, A\&A, 486, 177



\begin{table}

\begin{center}
\caption{Azimuthal velocity and metallicity from best fitting values for thin disc, thick disc and halo.}

\begin{tabular}{cccc}
 \hline\hline
 &$V_\phi$ ($\rm km\,s^{-1}$) & Metallicity (dex) \\

\hline
 Thin disc & $232.5^{+4.9}_{-7.9}$ & $-0.08^{+0.11}_{-0.07}$\\
Thick disc& $109.4^{+9.6}_{-2.9}$ & $-1.0^{+0.10}_{-0.12}$\\
Halo & $57.8^{+9.6}_{-12.9}$ & $-1.15^{+0.11}_{-0.13}$\\
\hline
\end{tabular}
\end{center}
\end{table}

\clearpage

\begin{figure}
\centering
\includegraphics[totalheight=2.6in,width=3.6in]{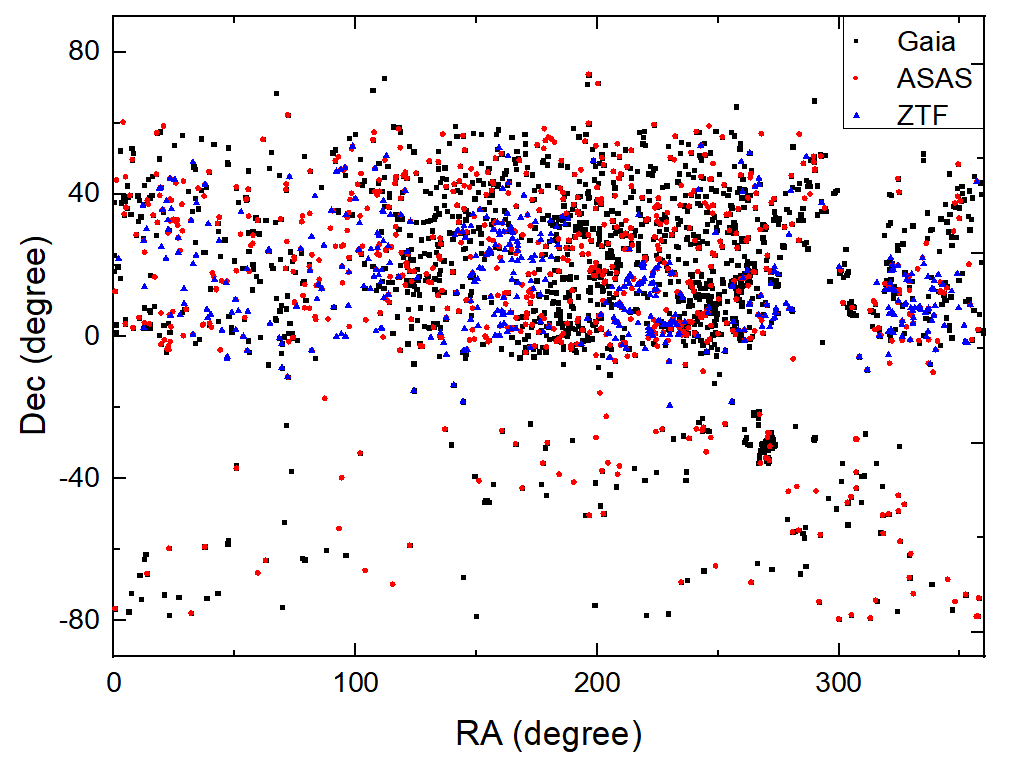}
\includegraphics[totalheight=2.7in,width=3.7in]{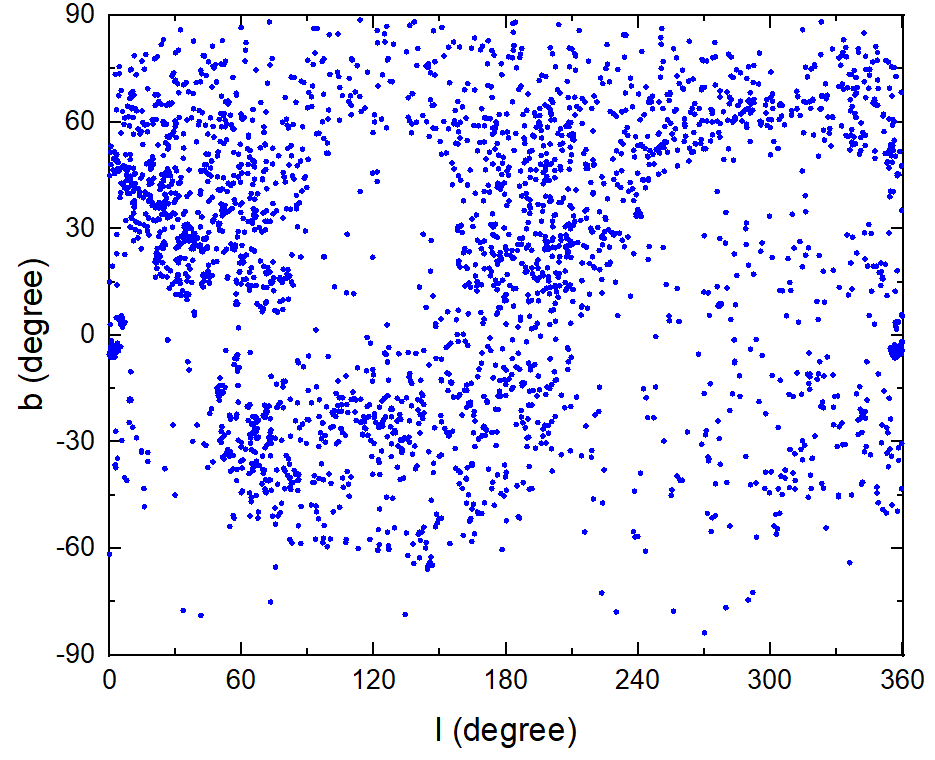}
\caption{The upper panel: Equatorial coordinates of RRLs identified from three surveys in our sample. The black squares are RRLs from Gaia catalog, red circles are RRLs from ASAS-SN catalog, and blue triangles are RRLs from ZTF catalog.
The lower panel: The distributions of the Galactic longitude ($l$) \& latitude ($b$) for all RRLs in our sample.}
\label{fig:1}
\end{figure}

\clearpage

\begin{figure}
\centering
\includegraphics[totalheight=4.9in,width=6.2in]{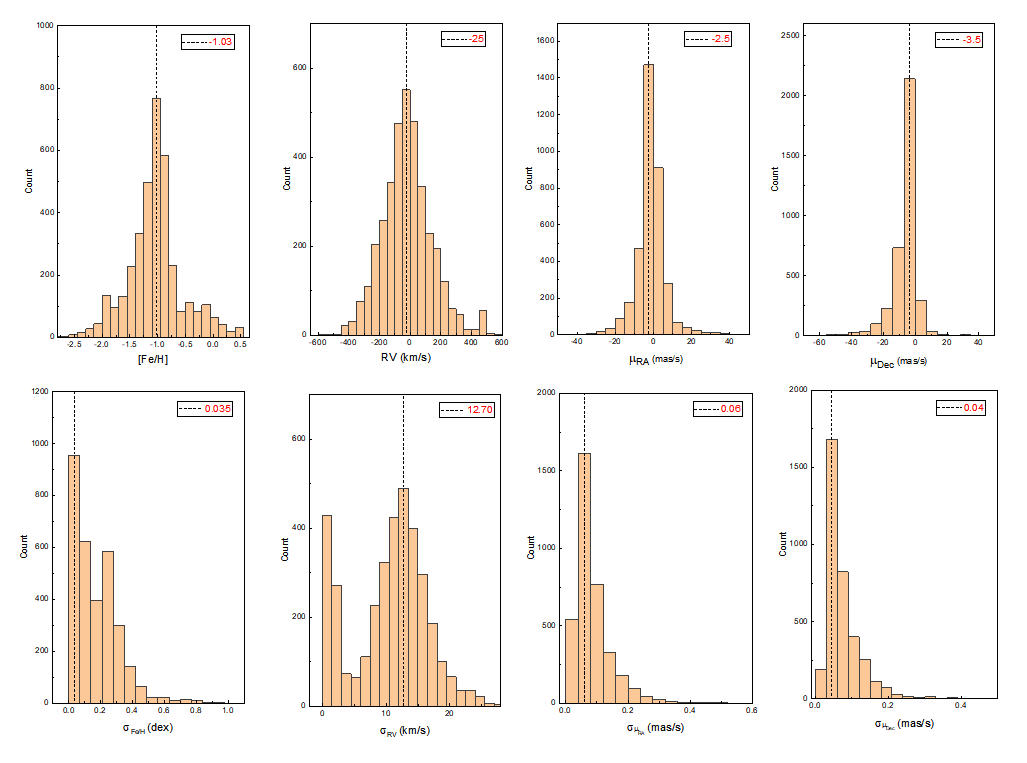}
\caption{The distributions of metallicity, radial velocity, proper motions are given in upper panels, while distributions of their uncertainties are shown in lower panels.
Dashed lines (red colored values) are the peak values in the distributions.}
\label{fig:1}
\end{figure}

\clearpage

\begin{figure}
\centering
\includegraphics[totalheight=5.0in,width=3.0in]{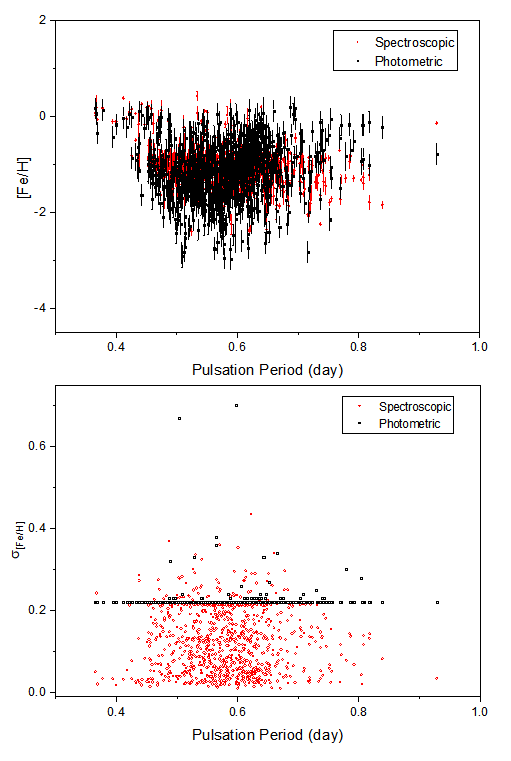}
\caption{Comparison of the photometrically derived metallicties (black squares with black lines (uncertainties)) and spectroscopically determined metallicities (red circles with red lines (uncertainties)) is given in the upper panel. The discrepancies between the uncertainties of them are shown in the lower panel. It is obvious that most of the spectroscopically determined metallicities have smaller uncertainties compared to photometrically derived metallicties (majority have uncertainties of 0.22 dex). }
\label{fig:1}
\end{figure}

\clearpage

\begin{figure}
\centering
\includegraphics[totalheight=2.3in,width=3.3in]{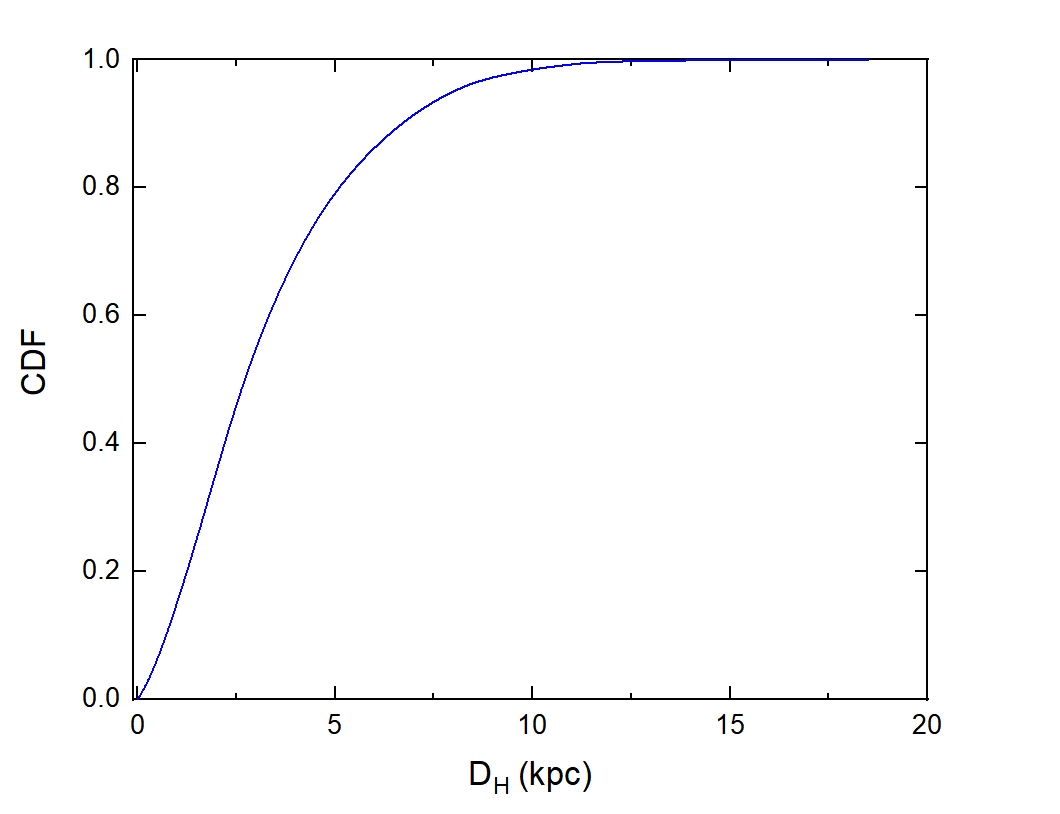}
\includegraphics[totalheight=2.4in,width=3.4in]{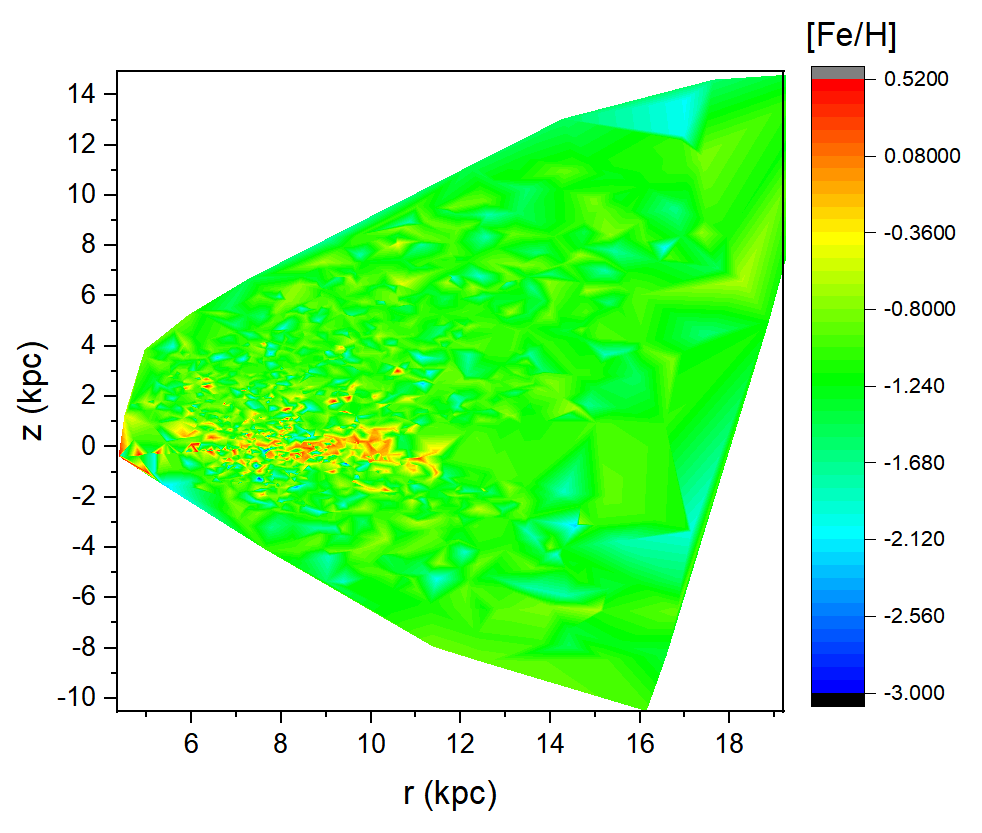}
\caption{Cumulative distribution function (CDF) of 3417 RRLs is given in the upper panel.
The lower panel is the color-map to show the metallicity distribution of the RRLs sample in the r and z plane. $r$ is the galactocentric distance in the Cartesian coordinate system, and $\rm z$ is the Galactic vertical height.}
\label{fig:1}
\end{figure}

\clearpage

\begin{figure}
\centering
\includegraphics[totalheight=4.5in,width=6.0in]{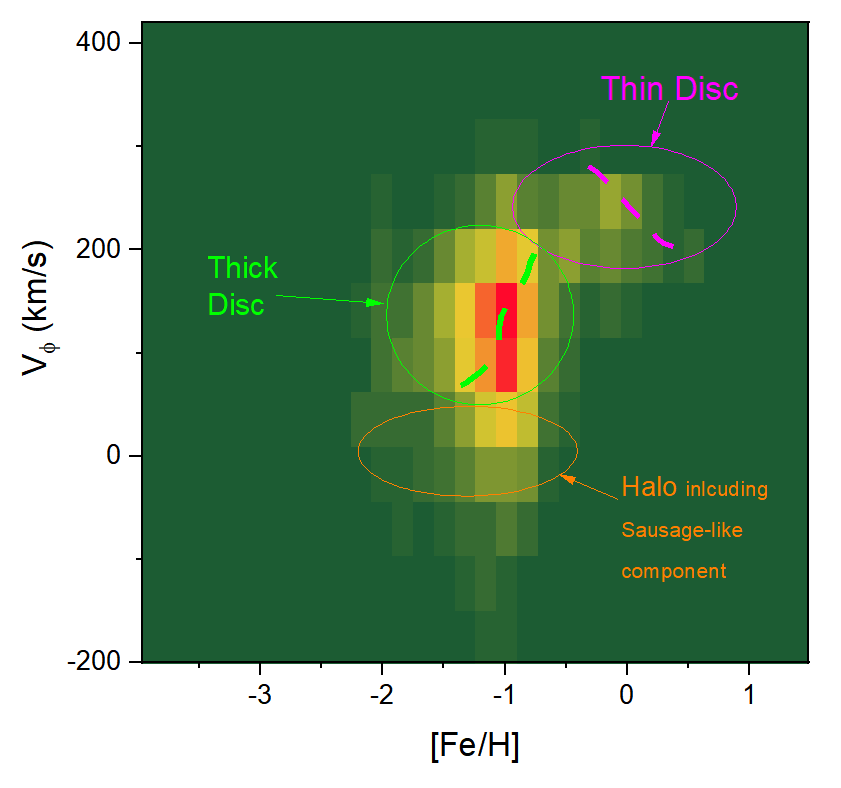}
\caption{Distribution of the azimuthal (in the direction of the Galactic rotation) velocity ($V_\phi$) versus metallicity ([Fe/H]). With the total 3417 RRL stars, it is weighted for each grid cell of the parameter space. The magenta area is the thin disc population, green area is the thick disc population, and orange area is the halo population including the GS (Sausage-like) population (see the text for more discussions).  }
\label{fig:1}
\end{figure}

\clearpage

\begin{figure}
\centering
\includegraphics[totalheight=4.5in,width=6.0in]{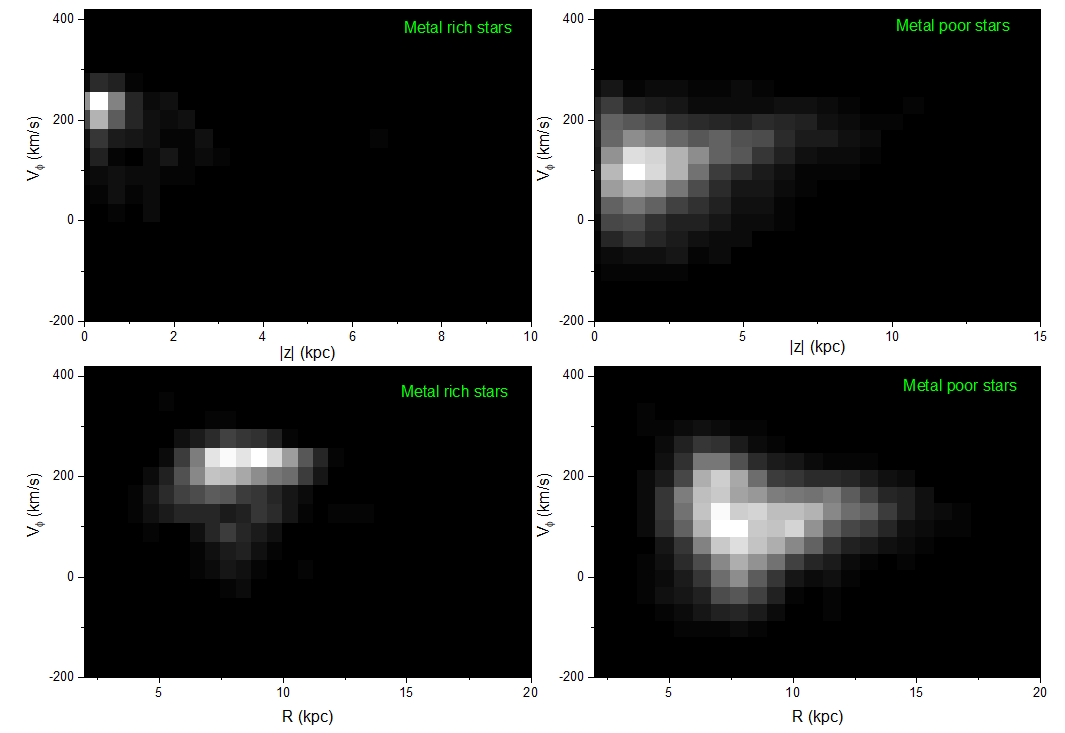}
\caption{Density maps for metal-poor and metal-rich stars in the parameter spaces of $V_\phi$ - Galactic vertical height $|z|$ (upper panels) and $V_\phi$ - Galactocentric distance in the Galactic plane $R$ (lower panels).
Here, RRLs with $[\rm Fe/H]>-0.7$ dex are metal-rich stars, and RRLs with $[\rm Fe/H]<-0.7$ dex are metal-poor stars.}
\label{fig:1}
\end{figure}

\clearpage

\begin{figure}
\centering
\includegraphics[totalheight=4.5in,width=6.0in]{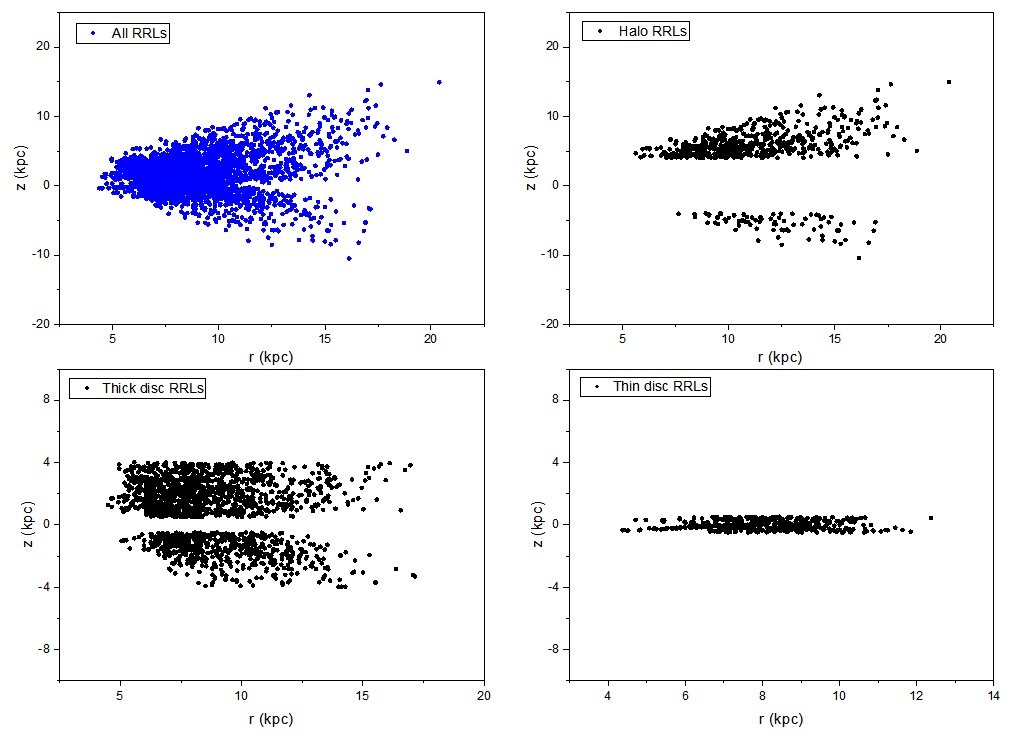}
\caption{Spatial distributions: all RRLs of the sample, halo RRLs, thick disc RRLs and thin disc RRLs in $r$ - $\rm z$ plane, respectively. }
\label{fig:1}
\end{figure}

\clearpage

\begin{figure}
\centering
\includegraphics[totalheight=4.5in,width=6.0in]{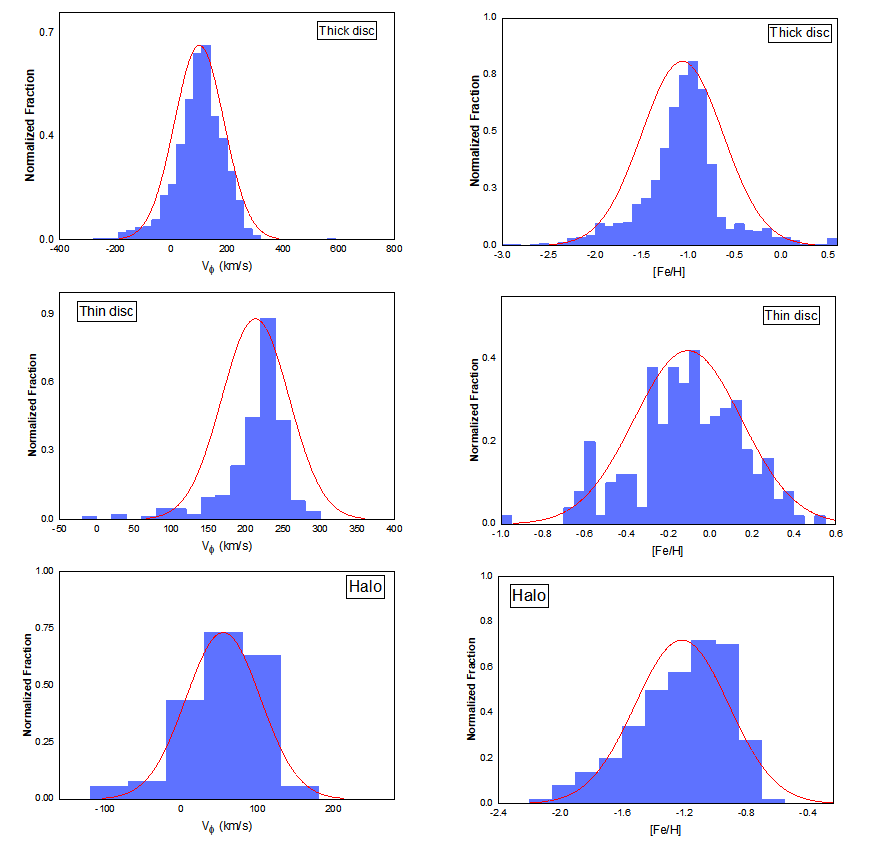}
\caption{Azimuthal Velocity and the metallicity distributions of RRLs selected for the thick disc, thin disc and halo, respectively. Red lines are the best fitted lines.}
\label{fig:1}
\end{figure}

\clearpage

\begin{figure}
\centering
\includegraphics[totalheight=2.3in,width=3.3in]{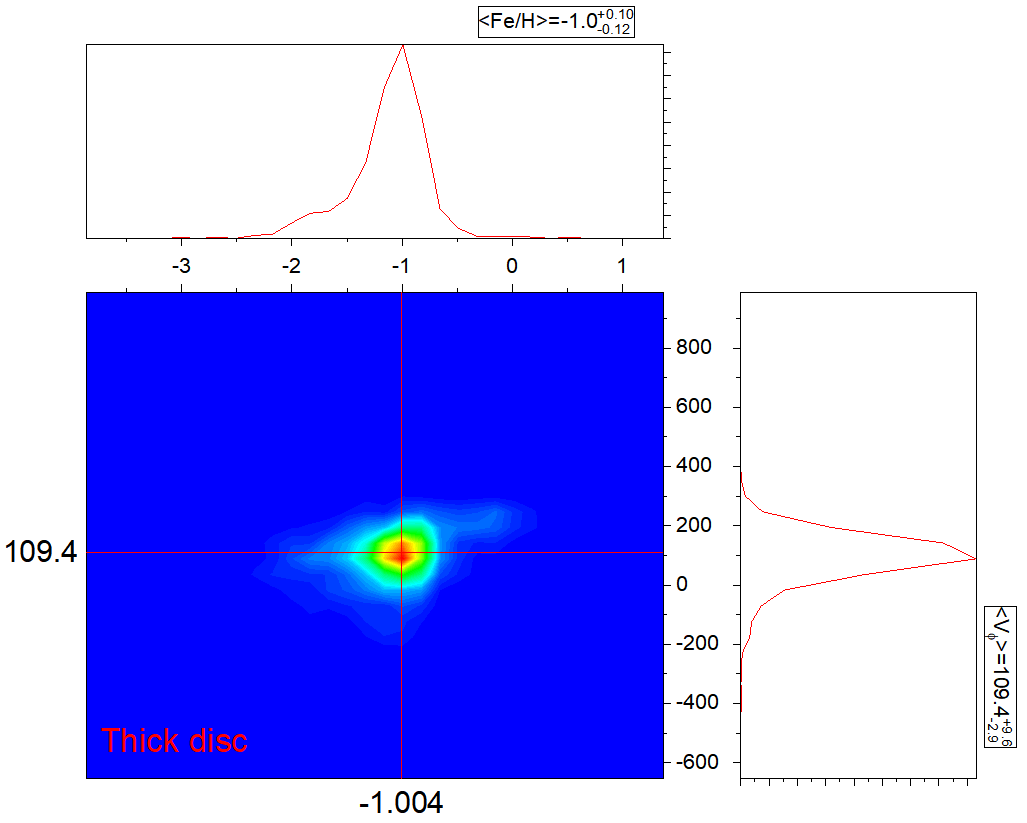}
\includegraphics[totalheight=2.3in,width=3.3in]{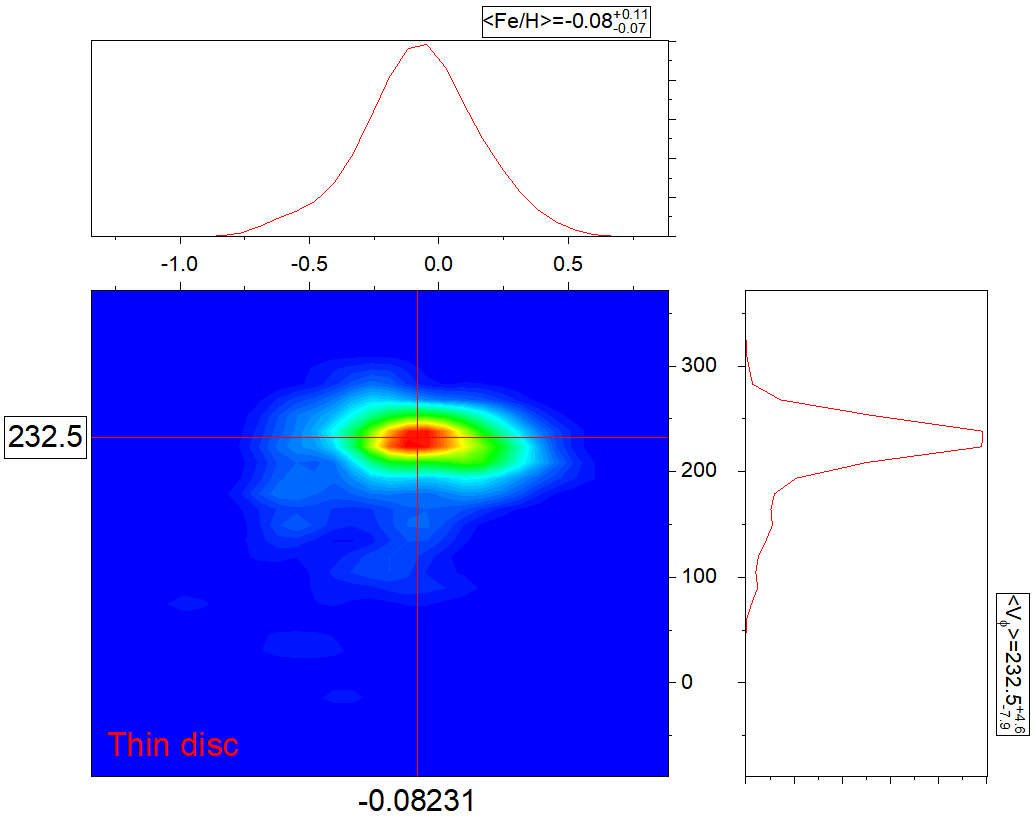}
\includegraphics[totalheight=2.3in,width=3.3in]{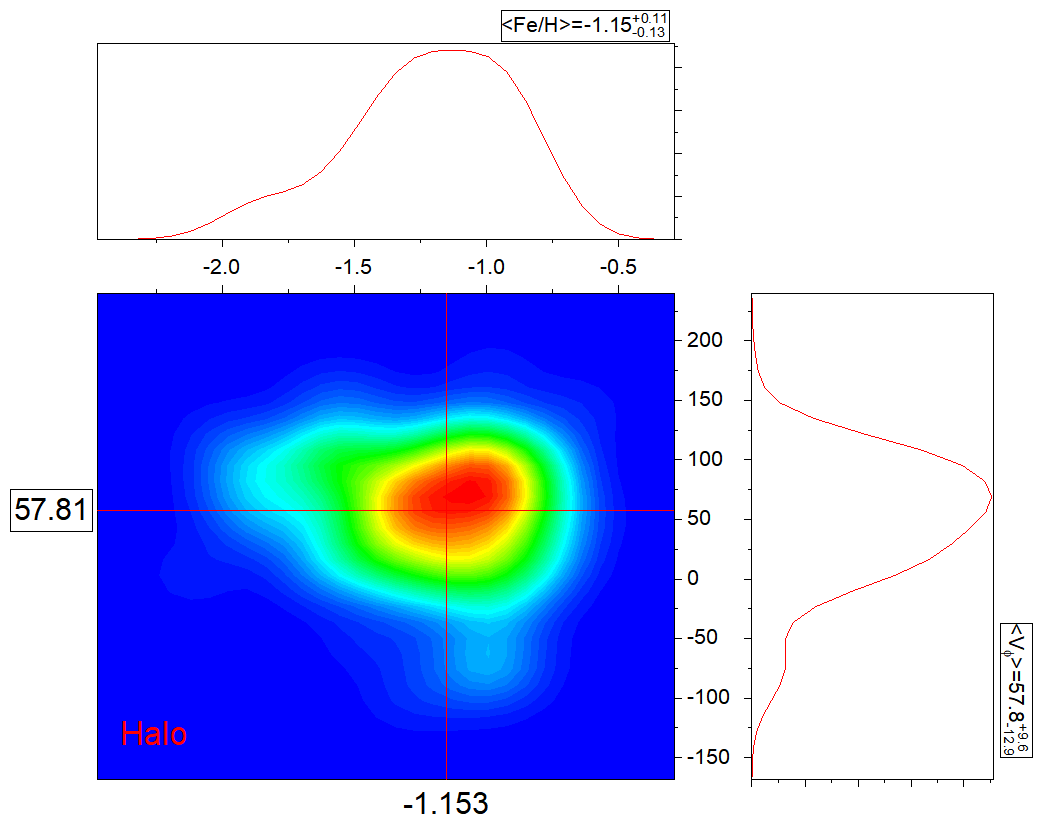}
\caption{Best fitted values from the Gaussian Model fitting results for three main components.}
\label{fig:1}
\end{figure}

\clearpage

\begin{figure}
\centering
\includegraphics[totalheight=4.5in,width=5.0in]{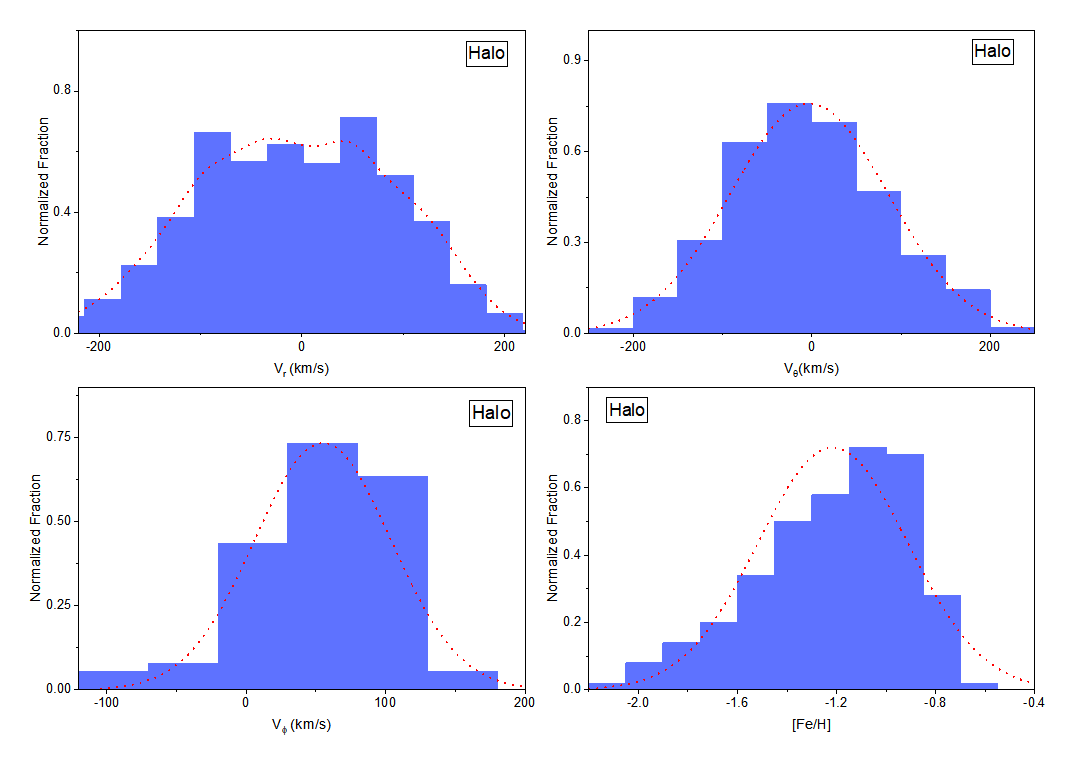}
\caption{Azimuthal Velocity and the metallicity distributions of the sample from the halo including the GS-like component.}
\label{fig:1}
\end{figure}

\clearpage

\begin{figure}
\centering
\includegraphics[totalheight=4.5in,width=5.0in]{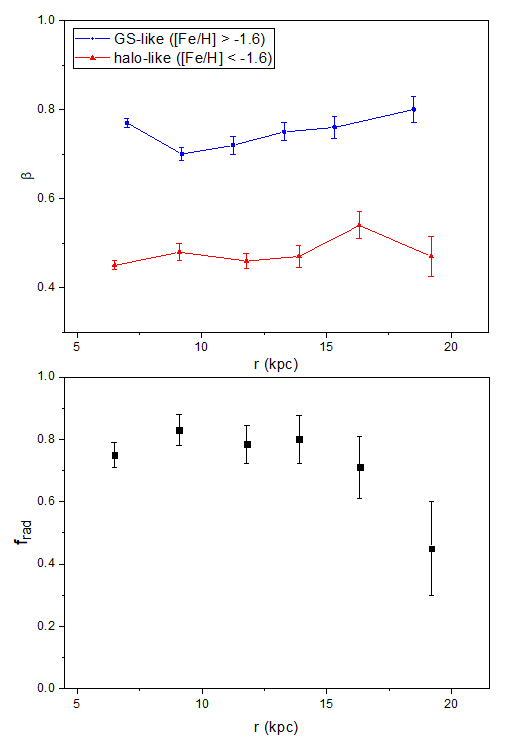}
\caption{Anisotropy of the GS-like (radial) component and its fraction in the total halo sample.}
\label{fig:1}
\end{figure}

\clearpage

\end{document}